\documentclass[final]{ias2}

\usepackage{graphicx} 
\usepackage{multirow}
\usepackage{array} 

\usepackage{hyperref} 

\begin{document}

\markboth{A New Solution of Einstein Vacuum Field Equations}{Ram Gopal Vishwakarma}

\newcommand{\be}{\begin{equation}}
\newcommand{\ee}{\end{equation}}
\def\bq{\begin{eqnarray}}
\def\eq{\end{eqnarray}}

\title{\bf A New Solution of Einstein's Vacuum Field Equations}
\author[ram]{Ram Gopal Vishwakarma}
\email{vishwa@uaz.edu.mx}
\address[ram]{Unidad Acad$\acute{e}$mica de Matem$\acute{a}$ticas,
 Universidad Aut$\acute{o}$noma de Zacatecas,
 C.P. 98068, Zacatecas, ZAC, Mexico}

\begin{abstract}
A new solution of Einstein's vacuum field equations is discovered which appears as a generalization of the well-known Ozsv$\acute{a}$th-Sch$\ddot{u}$cking solution and explains its source of curvature which has otherwise remained hidden.
Curiously, the new solution has a vanishing Kretschmann scalar and is singularity-free despite being curved.

The discovery of the new solution is facilitated by a new insight which reveals that it is always possible to define the source of curvature in a vacuum solution in terms of some dimensional parameters. As the parameters vanish, so does the curvature. 
The new insight also helps to make the vacuum solutions Machian. 
\end{abstract}



\maketitle

\section{Introduction}

Exact solutions of Einstein field equations have played important roles in the discussion of physical problems. Obvious examples are the Schwarzschild and Kerr solutions for  the study of black holes, and the Friedman solutions for cosmology. Of vital importance are the exact solutions of the vacuum field equations
\be
R^{\mu\nu}=0,\label{eq:RicciEq}
\ee
 which have tested general relativity (GR) beyond doubts and rendered it a well-established theory\footnote{The same is not true for the complete Einstein equations which require the speculative dark matter and dark energy in order to explain the cosmological observations. One the one hand, the theory predicts that about $27 \%$ of the total content of the Universe is made of non-baryonic dark matter particles, which should certainly be predicted by some extension of the Standard Model of particles physics. However, there is no indication of any new physics beyond the Standard Model which has been verified at the LHC. On the other hand, the dark energy, which according to the theory, constitutes about $68 \%$ of the total content of the Universe, poses serious confrontation with particle physics.}, 
since all the classical relativistic gravity effects predicted by GR are based on the solutions of equations (\ref{eq:RicciEq}) which have been successfully tested in the solar system and binary neutron stars.

The present article discovers a new solution of equations (\ref{eq:RicciEq}) which appears curious and explains the source of curvature of the well-known Ozsv$\acute{a}$th-Sch$\ddot{u}$cking (O-S) solution, which has remained an impenetrable mystery so far.
The motivation for the new solution ensues from a deeper insight into the theory which reveals that it is always possible to define the source of curvature in a vacuum solution in terms of dimensional parameters (which can support observable quantities such as the energy, momentum and angular momentum or their densities) present in the solution or in its variant.
These parameters appear in the Riemann tensor indicating source. As the parameters vanish, so does the tensor, reducing the solution to the Minkowskian form. 

We shall see in the following that this enlightened view of assigning the source to the dimensional constants, appears promising as  not only does it make definitive predictions about the new solutions of equations (\ref{eq:RicciEq}), which are indeed realized, but also helps to make the solutions Machian.

\section{A New Insight into the Source of Curvature in the Vacuum Solutions}

As gravitation is a manifestation of the curvature of spacetime with matter serving as the source of curvature, we should expect some source\footnote{Although other field equations, for example Maxwell's equations, 
possess source-free solutions and do not pose any consistency problem, however, the case of gravitation is different wherein the existence of a source-free solution goes against the requirements of Mach's principle.}
 of curvature in the curved solutions of equations (\ref{eq:RicciEq}). The source of curvature in a vacuum solution is conventionally assigned to a singularity appearing in the solution \cite{OS}. By this singularity we mean a true gravitational singularity (not a coordinate singularity which can be averted with a suitable choice of coordinates), which can be defined in a coordinate-independent way in terms of the divergence of the Kretschmann scalar\footnote{Though the presence of a singularity in a solution can be asserted unanimously by the divergence of the Kretschmann scalar $K$, the flatness of a solution cannot be asserted by the vanishing of $K$. For this we have to depend on the vanishing of the Riemann tensor. We shall come through some curved solutions for which $K$ will be vanishing!}
\be
K= R_{\alpha\beta\gamma\delta}R^{\alpha\beta\gamma\delta},
\ee
 where $R_{\alpha\beta\gamma\delta}$ is the Riemann tensor. However this conventional prescription of source in terms of singularity, does not seem to work universally as there also exist solutions of  equations (\ref{eq:RicciEq}), for example, the O-S solution and the Taub-NUT (Taub-Newman-Unti-Tamburino) solution, which are curved but singularity-free.
It appears that we have to define the presence of source in a smarter way which can be applied universally to all the solutions of equations (\ref{eq:RicciEq}).
In order to gain insight into the nature of the hidden source producing curvature in the solutions of equation (\ref{eq:RicciEq}), let us analyze critically its well-understood solutions.

\subsection{Schwarzschild Solution}

To begin with, let us first consider the Schwarzschild solution, which was discovered by Karl Schwarzschild in 1915 immediately after GR was formulated.
It is an exact solution of equations (\ref{eq:RicciEq}) and represents a static spacetime structure outside a static, isotropic mass, say $M$, placed in an empty space. In the Schwarzschild coordinates, the solution reads \cite{tolman}
\be
ds^2=\left(1+\frac{L}{r}\right)c^2 dt^2-\frac{dr^2}{(1+L/r)}-r^2d\theta^2-r^2\sin^2\theta ~d\phi^2,\label{eq:Sch}
\ee
where $L$ is a constant parameter appearing as a constant of integration in the solution. By requiring that GR should reduce to the Newtonian theory in the case of a weak
gravitational field, $L$ can be determined in terms of the source mass $M$ as
\be
L=-\frac{2GM}{c^2},\label{eq:K}
\ee
where $G$ and $c$ are respectively Newton's constant of gravitation and the speed of light in vacuum. Generally, the source of curvature in (\ref{eq:Sch}) is assigned to the singularity present at $r=0$, as is clear from the divergence of the Kretschmann scalar 
\be
K= \frac{12 L^2}{r^6}
\ee
at that point. Nevertheless, the presence of source can alternatively be defined by the presence of the `source carrier' parameter $L$. That the parameter $L$ contains the source,  is ascertained by the fact that it impregnates the Riemann tensor by appearing therein and as $L$ vanishes, so does the Riemann tensor reducing solution (\ref{eq:Sch}) to the Minkowskian form. 
Thus it is also possible to define the presence of source in solution (\ref{eq:Sch}) by the presence of the `source carrier' parameter $L$, without invoking the singularity.

\subsection{Kerr Solution}

If the mass $M$ rotates as well, the spacetime structure around it gets generalized with additional features and is given by the Kerr solution, which was discovered by Roy Kerr in 1963. If the mass is  spinning with an angular momentum per unit mass =  $\alpha$ (so that its total angular momentum $=Mc\alpha$), the solution (in the Boyer-Lindquist coordinates) takes the form \cite{hawking-ellis}
\begin{eqnarray}\nonumber
ds^2 &=& \left(1+\frac{Lr}{r^2+\alpha^2 \cos^2\theta}\right)c^2 dt^2-\left(\frac{r^2+\alpha^2 \cos^2\theta}{r^2+Lr+\alpha^2}\right)dr^2\\ \nonumber
&& -~(r^2+\alpha^2 \cos^2\theta)d\theta^2 -\left(r^2+\alpha^2-\frac{\alpha^2Lr}{r^2+\alpha^2 \cos^2\theta}\sin^2\theta\right)\\
&& \times \sin^2\theta ~d\phi^2-\left(\frac{2\alpha Lr}{r^2+\alpha^2 \cos^2\theta}\right)\sin^2\theta ~d\phi ~cdt,\label{eq:kerr}
\end{eqnarray}
for which the value of the Kretschmann scalar is obtained as
\be
K=\frac{12L^2[r^6-\alpha^2 \cos^2\theta(15 r^4-15\alpha^2 r^2 \cos^2\theta+\alpha^4 \cos^4\theta)]}{(r^2+\alpha^2 \cos^2\theta)^6}.\label{eq:K_kerr}
\ee
Obviously (\ref{eq:kerr}) is a generalization of the Schwarzschild solution (\ref{eq:Sch}) as it reduces to (\ref{eq:Sch}) for $\alpha=0$.
Equation (\ref{eq:K_kerr}) indicates the presence of a singularity at $(r=0, \theta=\pi/2)$ which is ascertained in a ring singularity by using the Kerr-Schild coordinates \cite{hawking-ellis}.
 
However, without invoking the singularity, it is readily possible to assign the source of curvature in solution (\ref{eq:kerr}) to the parameters $L$ and $\alpha$, which appear in the Riemann tensor, as we know. This shows that the angular momentum also contributes to the curvature, since the parameter $\alpha$ can be written in terms of the angular momentum, say  $J$, of the source mass:
\be
\alpha=\kappa_1\frac{J}{Mc},~~ (\kappa_1={\rm a~dimensionless~number}).
\ee
Vanishing of the Riemann tensor and hence reducing (\ref{eq:kerr}) to the Minkowskian form for vanishing $L$ and $\alpha$, asserts that these are the source-carrier parameters.

\subsection{Taub-NUT Solution}

This solution, which represents another generalization of the Schwarzschild solution, was discovered by Taub in 1951 \cite{Taub} and was rediscovered by Newman, Tamburino and Unti in 1963   \cite{NUT}. The solution can be written as 
\begin{eqnarray}\nonumber
ds^2 &=& \left(\frac{r^2+Lr-N^2}{r^2+N^2}\right)(c dt + 2N \cos \theta d\phi)^2 -\left(\frac{r^2+N^2}{r^2+Lr-N^2}\right) dr^2\\
&& -~(r^2+N^2)(d\theta^2+\sin^2\theta ~d\phi^2),\label{eq:Taub-NUT}
\end{eqnarray}
which reduces to the Schwarzschild solution for $N=0$. The solution does not possess any gravitational singularity and is perfectly regular at $r = 0$ as is indicated by its Kretschmann scalar
\begin{eqnarray}\nonumber
K &=& \frac{12}{(N^2+r^2)^6} [L^2r^6+N^2(4N^6-L^2N^4-24LN^4r-60N^4r^2\\ 
&& +15L^2N^2r^2+80LN^2r^3-15L^2r^4+60N^2r^4-24Lr^5-4r^6)].
\end{eqnarray}
Though it has the so-called `wire singularities' at $\theta = 0$ and $\theta =\pi$ where the metric
fails to be invertible, these are only coordinate singularities which can be removed by introducing two coordinate patches \cite{Misner}.

Thus being curved but singularity-free, solution (\ref{eq:Taub-NUT}) cannot explain the source of its curvature in terms of the conventional source singularities. Nevertheless, this situation can be averted if we define the source in terms of the `source carrier' parameters, here $L$ and $N$, which do appear in the Riemann tensor indicating source. While the parameter $L$ can be written in terms of the mass of the source (as we have seen in the Schwarzschild case), the parameter $N$ can be written in terms of the momentum $P$ of the source, i.e., 
\be
N=\kappa_2\frac{GP}{c^3},~~ (\kappa_2={\rm a~dimensionless~number}),
\ee
as can be checked by a simple dimensional analysis. Obviously the  parameters $L$ and $N$ are source-carriers as solution (\ref{eq:Taub-NUT}) becomes Minkowskian for $L=N=0$.

The novel representation of the source  in terms of the dimensional parameters, is also supported in the present case by the following observation: When (\ref{eq:Taub-NUT}) reduces to (\ref{eq:Sch}) for a vanishing $N$, we assert the presence of the source  (the point mass $M$) at $r=0$. This source must not disappear (with the disappearance of singularity) when the mass gains additional features (expressed through $N$) in  (\ref{eq:Taub-NUT}). This also reveals that the conventional representation of source through the singularities is not competent enough.

\subsection{Kasner Solution}

Another important solution of equations (\ref{eq:RicciEq}) is the Kasner solution, discovered by Edward Kasner in 1921 \cite{kasner}, and is given by 
\be
ds^2=c^2 dT^2- T^{2p_1}dX^2- T^{2p_2}dY^2- T^{2p_3}dZ^2,\label{eq:kasnero}
\ee
where the (dimensionless) parameters $p_1$, $p_2$ and $p_3$ satisfy
\[
p_1+p_2+p_3=1=p_1^2+p_2^2+p_3^2.
\]
The solution is interpreted in terms of an empty homogeneous Universe expanding and contracting anisotropically (for instance, for $p_1=p_2=2/3$ and $p_3=-1/3$, the space expands in two directions and contracts in the third). 
Although (\ref{eq:kasnero}) has a singularity at $T=0$, as is also witnessed by the Kretschmann scalar
\be
K=\frac{16(1-p_3)p_3^2}{T^4},
\ee
however, it does not seem to possess any dimensional parameter which could be associated to an observable quantity (like those which have been shown to source the curvature in the Schwarzschild, Kerr and Taub-NUT solutions) sourcing the curvature so that the solution can reduce to the Minkowskian form in the absence thereof. 
Though, solution (\ref{eq:kasnero}) can reduce to the Minkowskian form for $p_1=p_2=p_3=0$, these dimensionless pure numbers cannot support any of the above-mentioned observable quantities. However, a simple transformation 
\be
T=\frac{1+nt}{n}, ~X=n^{p_1}x, ~Y=n^{p_2}y, ~Z=n^{p_3}z, 
\ee
(where $n$ is an arbitrary constant parameter), can transform (\ref{eq:kasnero}) to a more useful form\footnote{The Kasner solution, in this form, was discovered by V. V. Narlikar and K. R. Karmarkar in 1946 \cite{curious}.}
\be
ds^2=c^2 dt^2- (1+nt)^{2p_1}dx^2- (1+nt)^{2p_2}dy^2
- (1+nt)^{2p_3}dz^2,\label{eq:kasner}
\ee
which serves the purpose. The non-vanishing components of the Riemann tensor for this metric yield 
\begin{eqnarray}\nonumber
&&R_{xyxy}=-n^2 p_1 p_2 (1+nt)^{-2p_3},  R_{xzxz}=-n^2 p_1 p_3 (1+nt)^{-2p_2},  \\\nonumber
&&R_{yzyz}=-n^2 p_2 p_3 (1+nt)^{-2p_1},  R_{xtxt}=n^2 p_2 p_3 (1+nt)^{-2(p_2+p_3)},\\
&&R_{ytyt}=n^2 p_1 p_3 (1+nt)^{-2(p_1+p_3)}, R_{ztzt}=n^2 p_1 p_2 (1+nt)^{-2(p_1+p_2)},
\end{eqnarray}
which contain $n$ actively such that they vanish for a vanishing $n$. A simple dimensional analysis suggests that the parameter $n$ can be written in terms of the momentum density, say ${\cal P}$, as
\be
n=\kappa_3 \sqrt{\frac{G{\cal P}}{c}},~~ (\kappa_3={\rm a~dimensionless~number}),\label{eq:en}
\ee
in order to meet its natural dimension in (\ref{eq:kasner}) (which is of the inverse of time).
As (\ref{eq:kasner}) reduces to the Minkowskian form for a vanishing ${\cal P}$,  the source of curvature present in (\ref{eq:kasner}), may be assigned to the net non-zero momentum density ${\cal P}$ arising from of the expanding/contracting spacetime given by (\ref{eq:kasner}) (even though the solution still has a singularity, as ascertained by the Kretschmann scalar 
\be
K=\frac{16n^4(1-p_3)p_3^2}{(1+nt)^4}, 
\ee
but now at $t=-1/n$ which is just a rescaling of time).

Although the representation of the source of curvature present in (\ref{eq:kasner}) in terms of the momentum density, does not make much sense in an empty space, nevertheless this interpretation is as good (or bad) as the standard interpretation of the expanding/contracting {\it empty} space without matter. If an utterly empty space without any matter can be conceived of expanding and contracting, it can very well  be associated with a momentum density. Moreover, the new interpretation provides useful clues to find a new solution of equations (\ref{eq:RicciEq}), as we shall see in the following.

\subsection{A Novel Representation of Source in Vacuum Solutions and its Predictions}

In the above, we could successfully interpret the source of curvature present in the solutions Schwarzschild, Kerr,  Taub-NUT and Kasner in terms of observable quantities contained in the dimensional constants $L$, $\alpha$, $N$ and $n$. which impregnate the Riemann tensor with their appearance. That the sources are contained in these parameters, is ascertained by their presence in Riemann tensor which vanishes as the parameters vanish, reducing the solutions Minkowskian.
This view is also supported by the Kerr-NUT solution which combines the Kerr and NUT solutions \cite{Dadhich}. Hence, it seems possible to interpret the source of curvature in a vacuum solution in terms of physical observable quantities sustained by some dimensional parameters present in the solution, {\it without taking recourse to the singularity}. This is also consistent with the de-Sitter solution 
\be
ds^2=c^2 dt^2- e^{2Ht} (dr^2+r^2d\theta^2+r^2\sin^2\theta ~d\phi^2),\label{eq:deSitter}
\ee
when we introduce the cosmological constant $\Lambda$ in the field equations\footnote{Although the cosmological constant (or any other candidate of dark energy) can very well be included in the energy-stress tensor, however the cosmological constant can also be a modification of the field equations and not necessarily a contribution to the energy-stress tensor, rather just a constant of nature.} 
and the curvature is generated by the scale of $\Lambda$, i.e., the constant $H=c\sqrt{\Lambda/3}$.

If this understanding is correct, it is expected to apply universally to all vacuum solutions, including the O-S solution (\ref{eq:OS}) (discussed later in section \ref{sec:OS}), which though does not appear to have any free parameter. The expectation is indeed met as we shall see in section \ref{sec:OS} that the O-S solution results from a more general solution by assigning a particular value to a free parameter of the solution.

The new insight appears promising as it makes definitive predictions:  if a solution of equations (\ref{eq:RicciEq}) can be supported by the momentum density as we noted in the case of the Kasner solution (\ref{eq:kasner}), we should also expect solutions of equations (\ref{eq:RicciEq})  supported by angular momentum density and energy density. In the following we consider the first prediction and discover that it is indeed realized in the form of a new vacuum solution sourced by the angular momentum density!

\section{A New Vacuum Solution}

Along the lines of the Kasner solution (\ref{eq:kasner}), let us expect a solution of equations (\ref{eq:RicciEq})  supported by the angular momentum density, say ${\cal J}$, (representing a rotating spacetime) such that the solution reduces to the Minkowskian form for a vanishing ${\cal J}$. In order to discover such a solution, we proceed as the following. 
By using ${\cal J}$, $G$ and $c$, one finds two possible ways to construct a parameter (of the dimensions of space or time, or their inverses), which vanishes when ${\cal J}$ vanishes. These are 
\begin{enumerate}
\item
~~~$\ell=\frac{G{\cal J}}{c^3}$, 

\item
~~~$m=\frac{G{\cal J}}{c^2}$, 
\end{enumerate}
which have the dimensions of the inverse of space and the inverse of time respectively. The first option appears suitable for a rotating spacetime, in which case the metric potentials are expected to be functions of the space-coordinates only. Tips from the Kasner metric (\ref{eq:kasner}) then help guessing the form of the metric representing a spacetime rotating about X-axis, which may be written, in a simple case,  as
\begin{eqnarray}\nonumber
ds^2 &=& (1+a_0 X^{b_0})c^2 dt^2 - (1+a_1 X^{b_1})dx^2
 - (1+a_2 X^{b_2})dy^2 \\\nonumber
&& -~(1+a_3 X^{b_3})dz^2 
+ a_{02}X^{b_{02}}cdt ~dy + a_{03}X^{b_{03}}cdt ~dz\\
&&  -~a_{12}X^{b_{12}}dx ~dy  - a_{23}X^{b_{23}}dy ~dz 
- a_{13}X^{b_{13}}dz ~dx,\label{eq:x}
\end{eqnarray}
where $X=\ell x$ and $a_i$, $b_i$, $a_{ij}$, $b_{ij}$ are constants to be determined by solving the field equations (\ref{eq:RicciEq}) for (\ref{eq:x}). The form of  the line element (\ref{eq:x}) is so chosen that it can reduce to the Minkowskian form for ${\cal J}=\ell=0$.
The absence of a term containing $dt dx$  asserts that (\ref{eq:x}) represents a spacetime rotating about X-axis. This tenet is supported by Adler, Bazin and Schiffer through the line element (4.83) in their book \cite{Adler}. Though, there can be more complicated forms of metric (\ref{eq:x}), let us start with this simple form. After making many `hits and trials' to solve (\ref{eq:RicciEq}) for $a_i$, $b_i$, $a_{ij}$, $b_{ij}$ of (\ref{eq:x}), finally we find the following solution
\begin{eqnarray}\nonumber
& a_1 = a_2=a_{12}=a_{13}=0, ~~a_3=-a_0=\frac{1}{8}, ~~a_{02}=a_{23}=1,~~a_{03}=\frac{1}{4}, \\
& b_0=b_3=b_{03}=2, ~~b_{02}=b_{23}=1,
\end{eqnarray}
which  renders  (\ref{eq:x}) in the following form, as an exact, curved solution of equations (\ref{eq:RicciEq}):
\begin{eqnarray}\nonumber
ds^2 &=&\left(1-\frac{\ell^2x^2}{8}\right)c^2dt^2-dx^2-dy^2-\left(1+\frac{\ell^2x^2}{8}\right)dz^2\\
&&+~\ell x (cdt-dz)dy+\frac{\ell^2x^2}{4}cdt~dz.\label{eq:int-kerr}
\end{eqnarray}
The non-vanishing components of the Christoffel symbol and the Riemann tensor, for this solution, are the following:
\begin{eqnarray}\nonumber
&& \Gamma^x_{ty}=\Gamma^t_{xy}=\Gamma^z_{xy}=\Gamma^y_{xz}=-\Gamma^y_{tx}=-\Gamma^x_{yz}
=\frac{\ell}{4},\\
&& \Gamma^x_{tz}=-\Gamma^x_{tt}=-\Gamma^x_{zz}=\frac{\ell^2 x}{8};
\end{eqnarray}
\be
R_{txtx}=R_{txxz}=R_{xzxz}=-R_{tyty}=-R_{tyyz}=-R_{yzyz}=\frac{\ell^2}{16}.\label{eq:Riem_i-k}
\ee
We note from (\ref{eq:Riem_i-k}) that $\ell$, and hence ${\cal J}$, does source the curvature of the new solution (\ref{eq:int-kerr}) which become flat for $\ell=0$, in consistence with our initial expectations. 
It may be curious to note that the Kretschmann scalar for (\ref{eq:int-kerr}) is vanishing, i.e. $K=0$ even for $\ell \neq 0$, though the solution is not Minkowskian then, as we can see from (\ref{eq:Riem_i-k})!
Remarkably, solution (\ref{eq:int-kerr}) is also singularity-free and strongly supports the new proposal of representing the source in terms of dimensional source-carrier parameters, here $\ell$.

\subsection{Ozsv$\acute{\rm a}$th-Sch$\ddot{\rm u}$cking Solution as a Particular Case of the New Solution (\ref{eq:int-kerr})}\label{sec:OS}

We have seen that solutions Schwarzschild, Kerr, Taub-NUT, Kasner and the new solution  (\ref{eq:int-kerr}), all contain dimensional `source carrier' parameters which can be attributed to the source of curvature. However, it does not appear so for the O-S solution which does not seem to carry any such parameters, neither a singularity.  In terms of coordinates $x^1, x^2, x^3, x^4$, the solution is given as 
\be
ds^2=-(dx^1)^2 +4x^4 dx^1 dx^3 -2 dx^2 dx^3 -2 (x^4)^2 (dx^3)^2 - (dx^4)^2.\label{eq:OS}
\ee
Discovered by Istv$\acute{a}$n Ozsv$\acute{a}$th and Engelbert Sch$\ddot{u}$cking in 1962 \cite{OS}, the line element  (\ref{eq:OS}) represents a stationary, geodesically complete and singularity-free solution of equations (\ref{eq:RicciEq}).   Interestingly, the Kretschmann scalar $K=0$ for this solution also, though it is curved as it has some non-vanishing components of the Riemann tensor given by $R_{3 4 3 4}=-R_{1 3 1 3}=1$. Hence its source of curvature has remained mysterious.
We shall see in the following that the O-S solution appears as a particular case of the new solution (\ref{eq:int-kerr}), which strengthens our belief in the new insight that the source of curvature in the vacuum solutions lies in the dimensional source-carrier parameters, and not in the singularities.

Let us write (\ref{eq:OS}) in new coordinates given by the transformation
\be
x^1=\bar{y}, x^2=-c\bar{t}-\bar{z}, x^3=c\bar{t}-\bar{z},x^4=\bar{x},
\ee
which transforms (\ref{eq:OS}) to a new form
\begin{eqnarray}\nonumber
ds^2 &=& 2c^2 (1-\bar{x}^2)d\bar{t}^2- d\bar{x}^2- d\bar{y}^2- 2 (1+\bar{x}^2)d\bar{z}^2+4c\bar{x}d\bar{t}d\bar{y}\\
&& +~4c\bar{x}^2d\bar{t}d\bar{z}-4\bar{x}d\bar{y}d\bar{z}.\label{eq:OSnew}
\end{eqnarray}
A further transformation
\be
\sqrt{2} \bar{t}=t,  \sqrt{2} \bar{z}=z, \bar{x}=x, \bar{y}=y
\ee
reduces solution (\ref{eq:OSnew}) to the form
\begin{eqnarray}\nonumber
ds^2 &=& (1-x^2)c^2 dt^2- dx^2- dy^2-(1+x^2)dz^2+2\sqrt{2}x(cdt-dz)dy\\
&&+~2x^2cdtdz,
\end{eqnarray}
which becomes a particular case of the new solution (\ref{eq:int-kerr}) for $\ell=2\sqrt{2}$, sourced by this particular value of the angular momentum density.
Hence the source of curvature in the O-S solution, in the absence of which the solution could resume the Minkowskian form, was not readily apparent since it had been assigned to this particular value. This also demonstrates that  the Ozsv$\acute{a}$th-Sch$\ddot{u}$cking solution may alternatively be interpreted as a rotating spacetime.

\section{Mach Principle and Vacuum Solutions}

As Mach's principle is shown to be well-supported by observations, there have always been attempts to derive Machian theories. Though many aspects and predictions of Mach's principle are realized in GR, the theory is not regarded completely Machian 
despite the fact that formulation of GR was facilitated by Mach's principle (refuting the Newtonian absolute space, {\it `which acts but cannot be acted upon'}).
This happens because GR admits solutions in which the principle supposedly does not hold. The vacuum solutions which are conventionally regarded to confront Mach's principle, have been the O-S solution and the Taub-NUT solution, and now  the new solution (\ref{eq:int-kerr}) also added to the list. 
However, the success of GR at various theoretical and observational fronts and the consistency of GR with special relativity (which too abolishes the absolute space akin to Mach's principle) cry out with all force that the theory must be Machian. 

Among various statements of Mach's principle available in GR, there is one version of this postulate given by Pirani \cite{OS} that has been defined unambiguously: ``In the absence of source matter, spacetime should necessarily be Minkowskian".  If the presence of source is defined by the presence of (true gravitational ) singularities in the vacuum solutions (as is in practice \cite{OS}), the principle would forbid a singularity-free vacuum solution from being anything but Minkowskian.
Hence the principle is violated in the solutions  O-S, Taub-NUT and the new solution (\ref{eq:int-kerr}) which are curved but singularity-free, as we have seen. This perhaps insinuates that we have to define the presence of source in a smarter way which can apply universally to all the solutions of equations (\ref{eq:RicciEq}), and not through the singularity which is anyway an unphysical feature and a sign of breakdown of the theory.

We have already developed a new insight into the underlying source of curvature in the vacuum solutions which reveals that it is always possible to attribute the source of curvature in a vacuum solution to one or more dimensional parameters, without taking recourse to the singularities;  as the parameters vanish, so does the curvature. These parameters can support physical observable quantities such as the energy, momentum and angular momentum or their densities. It appears that if the presence of source in the vacuum solutions is defined by the presence of such parameters, as should reasonably be expected, all the solutions become Machian!
Moreover, it provides a new meaning to the field equations (\ref{eq:RicciEq}) \cite{frontier}.

This enlightened view of locating Mach's principle in the source of curvature through the source-carrier parameters, appears rewarding as  not only does it solve a foundational problem, but also provides a better understanding of the theory.

\section{Conclusion}

A new insight into the source of curvature in vacuum solutions is developed which reveals that it is always possible to identify the source of curvature with some dimensional parameters present in the solutions;  as the parameters vanish, so does the curvature. This new proposal of defining the source of curvature in terms of these source-carrier parameters, is important in its own right, as it presents a new scope to represent the curvature in the vacuum solutions which apply universally to all vacuum solutions. More importantly, the new proposal facilitates to discover a new solution of Einstein vacuum field equations which is curved but singularity-free. The solution strongly supports the novel prescription of defining source in terms of the source-carrier parameters and displays the incompetence of the conventional prescription of defining source in terms of the singularities. The new solution also explains the source of curvature of the well-known Ozsv$\acute{a}$th-Sch$\ddot{u}$cking solution, which has otherwise remained hidden.

Interestingly, all the vacuum solutions become Machian in the new definition of the presence of source given by the presence of the `source carrier' parameters, remedying the unphysical, lingering situation. It may be noted that the solutions Tab-NUT, Ozsv$\acute{a}$th-Sch$\ddot{u}$cking and the new solution discovered in the paper, which are curved but singularity-free, contradict Mach's principle if the presence of source matter is defined by the conventional prescription of the presence of singularity in the vacuum solutions.

\bigskip

\noindent
{\bf Acknowledgment:} The author thanks Vijay Rai for sending some old, important literature and IUCAA for hospitality where part of this work was done.

\bigskip

\bibliographystyle{pramana}
\bibliography{references}

\end{document}